\documentclass[aps,prb,reprint,showpacs,floatfix]{revtex4-1}
\pdfoutput=1
\usepackage{balance}  
\usepackage{graphicx}
\usepackage{amsmath}
\usepackage{amsthm}
\usepackage[normalem]{ulem} 
\usepackage{subfigure}
\usepackage{empheq}

\newtheorem{thm}{Theorem}
\newtheorem{cnj}{Conjecture}
\newtheorem{definit}{Definition}
\newcommand{\iu}{\mathrm{i}\mkern1mu}
\newcommand{\suchthat}{\mathrel{\mathop\supset}\kern-4.0pt$-$\kern-1.0pt$-~$}

\begin{document}
\preprint{Medical Physics}
\title[Stable distributions for Proton Beams]{Using Stable Distributions to Characterize Proton Pencil Beams}
\author{Frank Van den Heuvel}
\affiliation{CRUK/MRC Oxford Institute for Radiation Oncology,\\
University of Oxford, UK}
\email[Email: ]{frank.vandenheuvel@oncology.ox.ac.uk}
\thanks{Author to whom correspondence should be addressed.}
\author{Francesca Fiorini }
\affiliation{CRUK/MRC Oxford Institute for Radiation Oncology,\\ University of Oxford, UK}
\author{Niek Schreuder}
\affiliation{Department of Medical Physics,
Provision Center for Proton Therapy,
Knoxville, Tennessee}
\author{Ben George} 
\affiliation{CRUK/MRC Oxford Institute for Radiation Oncology,\\ University of Oxford, UK}
\begin{abstract}
{\bf Purpose:}
To introduce and evaluate the use of stable distributions as a means of describing the behavior of charged particle pencil beams in a medium, with specific emphasis on proton beam scanning (PBS).
\par
{\bf Methods:}
The proton pencil beams of a clinically commissioned proton treatment facility are replicated in a Monte Carlo simulation system (FLUKA). For each available energy the beam deposition in water medium is characterized by the dose deposition. Using an alpha--stable distribution methodology each beam with a nominal energy $E$ is characterized by the lateral spread at depth $z$:  $S(z;\alpha,\gamma,E)$ and a total energy deposition $I_D(z)$.
The beams are then described as a function of the variation of the parameters at depth. Finally, an implementation  in a freely available open source dose calculation suite (matRad, DKFZ, Heidelberg, Germany) is proposed. 
\par 
{\bf Results:} 
Quantitatively, the fit of the stable distributions, compared to those implemented in standard treatment planning systems, are equivalent. The efficiency of the representation is better (2 compared to 3 and more parameters needed). The meta--parametrization (i.e. the description of the dose deposition by only providing the fitted parameters) allows for interpolation of non--measured data. In the case of the clinical data used in this paper, it was possible to only commission 1 out of 5 nominal energies to obtain a viable data set.  
\par
{\bf Conclusions:}
Alpha--stable distributions are intrinsically suited to describe charged particle pencil beams in a medium and can be easily implemented in existing treatment planning systems. The use of alpha-distributions can easily be extended to other particles.
\end{abstract}
\maketitle

\section{Introduction}
In proton beam therapy treatment planning, analytical descriptions of the treatment beams are commonly used to determine the dose deposited in clinical patient models. Although Monte Carlo based methods have become faster during the last few years, there still is a distinct advantage of using more efficient, analytical models when having to perform multiple calculations such as in the process of 4D robust--optimization and adaptive therapy. This advantage has greater significance in the case of pencil beam based proton therapy, where multiple small beams need to be tracked and calculated. Despite all these advantages, analytical algorithms have been shown to be less reliable in clinically more complex treatments like lung and breast treatments\cite{fiorini2015}, which prompted this effort to provide a more accurate description of the dose deposition by a scanned proton beam.  
In addition, a more analytical description provides greater insight in the macroscopic process of how a pencil beam behaves physically in a medium as issues such as energy and medium vary. 

\par
\begin{figure}[h!]
\centering
\includegraphics[width=0.7\columnwidth]{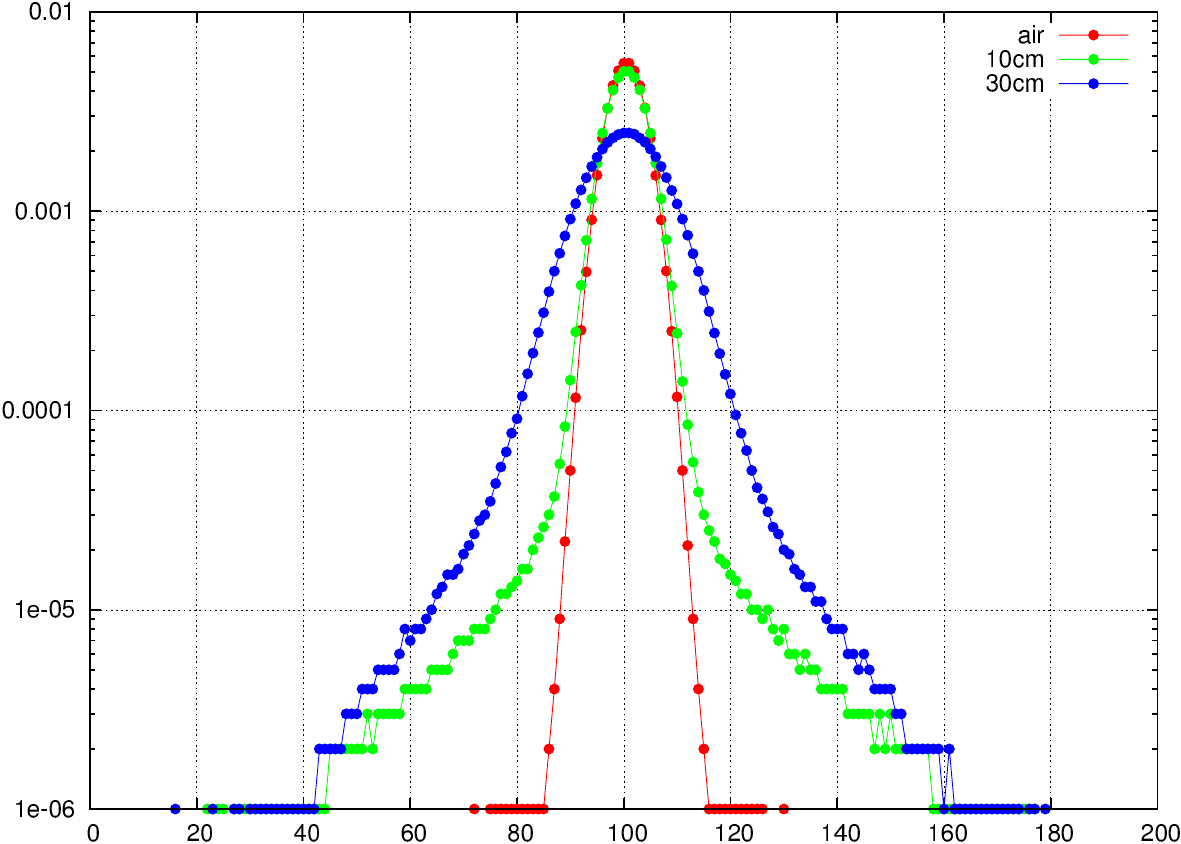}
\caption{\label{depth_change}Monte Carlo--based calculation showing changes in lateral dose deposition for a pencil beam of nominal energy of 230 MeV. A logarithmic scale is used to better illustrate the difference in contributions from the nuclear halo and the primary particles.}
\end{figure}

A pencil beam entering a medium will generate secondary particles such as scattered neutrons, generated photons, $\delta$--rays and large angle scattered protons that produce a {\em nuclear halo} of dose around the central beam axis. Although, in the region away from the central axis, the contribution from a single pencil beam is small, a complex treatment plan is made of many pencil beams and the summation of the lateral contributions could be significant. The lateral extent of a beam at different depths is illustrated in Figure \ref{depth_change}.
\par
The description and treatment of the nuclear halo has been the focus of research by a number of groups who have proposed various methodologies describing the effects in an analytical way. Gottschalk et al. provided an in-depth analysis of all the physical processes contributing to the nuclear halo, subdividing a pencil beam into a combination of four distinct regions; core, halo, aura and (possibly) spray; in the pencil beam\cite{Gottschalkarx2014,Gottschalk2015}. This approach requires up to 25 different physical parameters to characterise the beam. In most implementations, the contribution from the nuclear halo is solved by adding different distributions to a central Gaussian distribution describing the core of the pencil beam.
\par
The first proposed solution for the nuclear halo by Pedroni et al. added another, broader Gaussian to the core; a methodology that is implemented in the Varian Eclipse (Varian Medical Systems, Palo Alto, CA) calculation algorithm\cite{pedroni2005}. The simplicity of this calculation also allowed for faster GPU implementation\cite{daSilva2015.ROHTUA}. In further refinements of this approach, other groups attempted combinations of Gaussian, Lorentz (also known as Cauchy) and L\'evy distributions\cite{Bellinzona2015}, increasing the complexity of the fitting procedure and necessitating look-up tables for the various parameters. A key insight that enables our novel approach is that each of these methods combine two or more stable functions in their analytical representation. A further  clinically used algorithm for pencil beam calculation is used in RayStation\texttrademark TPS (RaySearch Laboratories, Stockholm, Sweden). In this system, each spot is modelled as a superposition of 19 Gaussian distributions (19 sub-spots: 1 at the center, and 6 and 12 positioned at two concentric circles around the center)\cite{RSmanual}.
\par
In this paper we review the concept of stable distributions and show that they can be used to represent the evolution of a proton pencil beam in a medium. We demonstrate that this approach provides a more accurate description of the pencil beam and is more efficient than the use of normal distributions, or a sum thereof. Furthermore, we show that this parametrization allows interpolation of non--measured energies from measured (or calculated using Monte Carlo) depth profiles. Finally, we implement this algorithm in an open source treatment planning toolkit, matRad\cite{cisternas2015matrad}.

\section{Methods and materials}
\subsection{Stable distributions}
Stable distributions are a class of distributions which generalize a property of the normal distribution. Namely, they extend the central limit theorem which says that if the number of samples drawn from random variables, \emph{with or without} finite variance, tends to infinity, then the measured distribution tends to a stable distribution. If the variance is finite, the resultant distribution tends towards the normal distribution, a member of the class of stable distributions.

\par
Other than for specific cases, these distributions do not possess an analytical representation. It is therefore necessary to describe them in terms of their characteristic function which always exists for a given stable distribution. 

\par
More generally, the characteristic function, $\varphi(t)$, of a distribution is the Fourier transform of the probability function, $f(x)$, of that distribution, e.g.:
\begin{equation}
\varphi(t) = \frac{1}{2\pi}\int_{-\infty}^{\infty} f(x)e^{-\iu xt}dx
\end{equation}
It can be shown that all stable distributions can be characterised as having the same characteristic function, $\varphi(t)$, barring a change in the parameters $(\alpha,\beta,\gamma,\delta)$.
\begin{equation}
\label{cf_long}
\varphi(t;\alpha,\beta,\gamma,\delta)~=~\exp [\iu t\delta - |\gamma t|^\alpha(1-\iu\beta \mathrm{sgn}(t) \phi)]
\end{equation}
With $\phi(t) = \tan(\pi\alpha/2)$ except for $\alpha = 1$, in which case $\phi(t)=-\frac{2}{\pi}\log(t)$. The parameter $\alpha \in [0,2]$ determines the shape of the distribution, $\beta \in [-1,1]$ is a measure for symmetry, $\gamma \in [0,+\infty]$ is a scale factor and $\delta$ a position, or the most probable value\cite{1999_stable}. For a symmetric, zero centred distribution the Equation \ref{cf_long} reduces to:
\begin{equation}
\label{cf_short}
\varphi(t;\alpha,\gamma)~=~\exp(- |\gamma t|^\alpha)
\end{equation}
In an appendix we show from first principles that this equation represents all symmetric zero--centered  distributions that follow the central limit theorem. 
\par
As $\alpha$ and $\gamma$ can vary continuously there are an infinite number of stable distributions, most of which do not have an analytical representation in real space. Indeed, only  for $\alpha = 2$, $1$, and $0.5$ ($\beta = 1$) is a closed form known. These correspond to, respectively, the Gauss--,  Lorenz-- , and L\'evy--distributions.

\par
Using this generalization it is possible to define a class of uni--modal distributions whose properties can be exploited to describe physical random walk processes which combine different physical properties\cite{Humphries2010}.

\subsection{Monte Carlo Simulations}
The ProVision Center for Proton Therapy is currently operational using an IBA cyclotron which provides proton beam scanning technique up to a maximum energy of 230 MeV. For this work, the ProVision beams from 98 to 230 MeV were accurately reproduced by the Monte Carlo code FLUKA by adapting the simulated beams to the commissioning experimental beam data\cite{fluka1,fluka2}. At each energy, the beam is defined at the surface of the phantom by a two-dimensional normal distribution characterized by position and standard deviation, $\sigma$. Using FLUKA, the dose distribution in medium (water) is calculated in a 200x200x350 mm$^3$ cube with 1 mm$^3$ tally volumes. The calculated dose distribution in each 200x200 slice perpendicular to the beam axis is then considered to be a two--dimensional dose distribution. Generation of secondary particles under the form of gamma's, neutrons and $\delta$--rays was enabled.

\subsection{Comparative algorithms}
The simplest approach of analytically predicting the behavior of the
dose deposited by a proton beam is based on a Gaussian parametrization
with the width of the beam defined by a variable standard deviation,
$\sigma$\cite{Gottschalk2015}. This implies that at any given depth the
lateral dose deposition can be described as:
\begin{equation}
D(z,E) = \frac{1}{2\pi \sigma^2\left(z,E\right)}\exp\left(-\frac{r^2}{2\sigma^2\left(z,E\right)}\right)
\end{equation}

In commercially available algorithms it was seen that this approach did
not predict the contributions of pencil beams further away from the beam
axis in an adequate way. Indeed in the  Eclipse\texttrademark treatment
planning software, developed by Varian (Varian Medical Systems, Palo Alto,
CA), an additional Gaussian term is used to describe this, as proposed by Pedroni {\it et al.}\cite{pedroni2005}. RayStation (Raysearch, Stockholm Sweden)
also employs such a strategy introducing multiple off axis Gaussian contributions.
In the remainder of the paper we will only concentrate on the two Gaussian
solution, which is parametrized as follows:
\begin{align}
D(z,E) &= \frac{q}{2\pi \sigma_1^2\left(z,E\right)}\exp\left(-\frac{r^2}{2\sigma_1^2\left(z,E\right)}\right) \\
&+ \frac{\left(1-q\right)}{2\pi \sigma_2^2\left(z,E\right)}\exp\left(-\frac{r^2}{2\sigma_2^2\left(z,E\right)}\right)
\end{align}
with $q \in [0,1]$. This parametrization implies that to fit the behaviour of a proton pencil beam at a given depth we need to determine three parameters: $\sigma_1$, $\sigma_2$, and $q$.

\subsection{Fitting procedures}
Because the majority of stable distributions do not possess an
analytic form; it is difficult to use the classical approach
to fit the data. Indeed, the fitting of stable distributions
is the subject of scientific research by itself. We opted to
use a maximum likelihood estimation based on pre--computed spline
approximations\cite{Nolan2001}. In essence it selects the distributions
that match the pre--computed ones the best.
\par
Once the parameters are determined, the characteristic function is
calculated in complex space and using an inverse Fourier transform
the actual stable distribution was generated, a straightforward
methodology also proposed by Mittnik et al.\cite{Mittnick1999}.

The resulting curve  could then be compared with the Monte Carlo simulation.
\par
To fit the normal distribution based algorithms, a classical
methodology using a least square fit of the analytical function based
on a Levenberg--Marquardt algorithm\cite{Gnuplot_4.4} was used.
\par
Both the stable and Gaussian fit were compared to the Monte Carlo simulation using Pearson's $\chi^2$ measure.
\begin{equation}
\chi^2 ~=~ \sum_{i=1}^N \frac{(E_i -O_i)^2}{E_i}
\end{equation}
The $\chi^2$ value for $200-n$ degrees of freedom then yields the probability
that the fitted distribution is different from the simulated one, with $n$ being the number of parameters in the fit. For stable parametrization $n = 2$ ($\alpha$ and $\gamma$), for a double Gaussian $n = 3$ ($\sigma_1$, $\sigma_1$, and $q$).  
We denote the $\chi^2$ value for
normal and stable distribution as respectively $\chi^2_N$ and $\chi^2_S$.

\subsection{Parametrization and scaling}
For all pencil beams with nominal energy ($E_N$) the centrally located transversal
distribution is extracted at all available depths ($z$). Subsequently, the
normalised stable distribution parameters $\alpha(z,E_N)$ and $\gamma(z,E_N)$
are determined using the above-mentioned fitting procedure. In a
first approximation we consider the pencil beams to be circularly--symmetric. Finally, the total integral dose at each depth $D(z,E_N)$ is
also calculated. This procedure yields three parameters which vary as
a function of depth and nominal beam energy allowing us to calculate the dose distribution at any depth in a homogeneous medium. 

\subsection{Data Interpolation}
We propose a methodology to determine the beam characteristics of intermediate energies from two provided beam characterisations. 
\begin{cnj}[Intermediate Morphing]
Let $\alpha(z,E_i)$, $\gamma(z,E_i)$, and $D(z,E_i)$ be the parameters
fully describing a proton pencil beam with nominal energy $E_i$. Then it is possible to calculate the parametrization of an intermediate energy $E_j$ by interpolation of the parametrization of energies $E_i$ and $E_k$ disregarding a scaling in the depth parameter which depends on the range of the given energy.
\newline 
let $E_i<E_j<E_k$, :
\begin{align}
\alpha(z',E_j) &= \alpha(z,E_i) + \frac{E_j -E_i}{E_k - E_i}(\alpha(z'',E_k) - \alpha(z,E_i))\\
\gamma(z',E_j) &= \gamma(z,E_i) + \frac{E_j -E_i}{E_k - E_i}(\gamma(z'',E_k) - \gamma(z,E_i))\\
D(z',E_j) &= D(z,E_i) + \frac{E_j -E_1}{E_k - E_i}(D(z'',E_k) - D(z,E_i))
\end{align}
where:
$z' = \Re(E_1)/\Re(E_2)$ and $z'' = \Re(E_1)/\Re(E_3)$ with $\Re(E)$ being the range of a proton in the medium under consideration.

\end{cnj}

Using the methodology of intermediate morphing, we determine the minimal amount of beams we need to fully characterize in order to obtain a full set of data across all nominal energies. We choose a threshold of 1\% error determining the width of the beam ($\gamma$) and 3\% for the shape (tailedness) ($\alpha$). The total deposited energy needs to be correct to 1\% dose and 1 mm position. 

\subsection{Implementation in matRad}
To allow testing of our parametrised beam model with clinical patient plans, we implemented the stable distribution dose calculation algorithm in an open source treatment planning system, matRad (DKFZ, Heidelberg, DE). matRad is written in MATLAB (MathWorks, Natick, MA) and provides functionality for importing patient data, ray tracing, inverse planning and treatment plan visualisation. The proton dose calculation component was extended to support a beam model described by a stable distribution, in addition to single and double Gaussian models.

\par
\begin{figure}[h!]
\centering
\includegraphics[width=1.0\columnwidth]{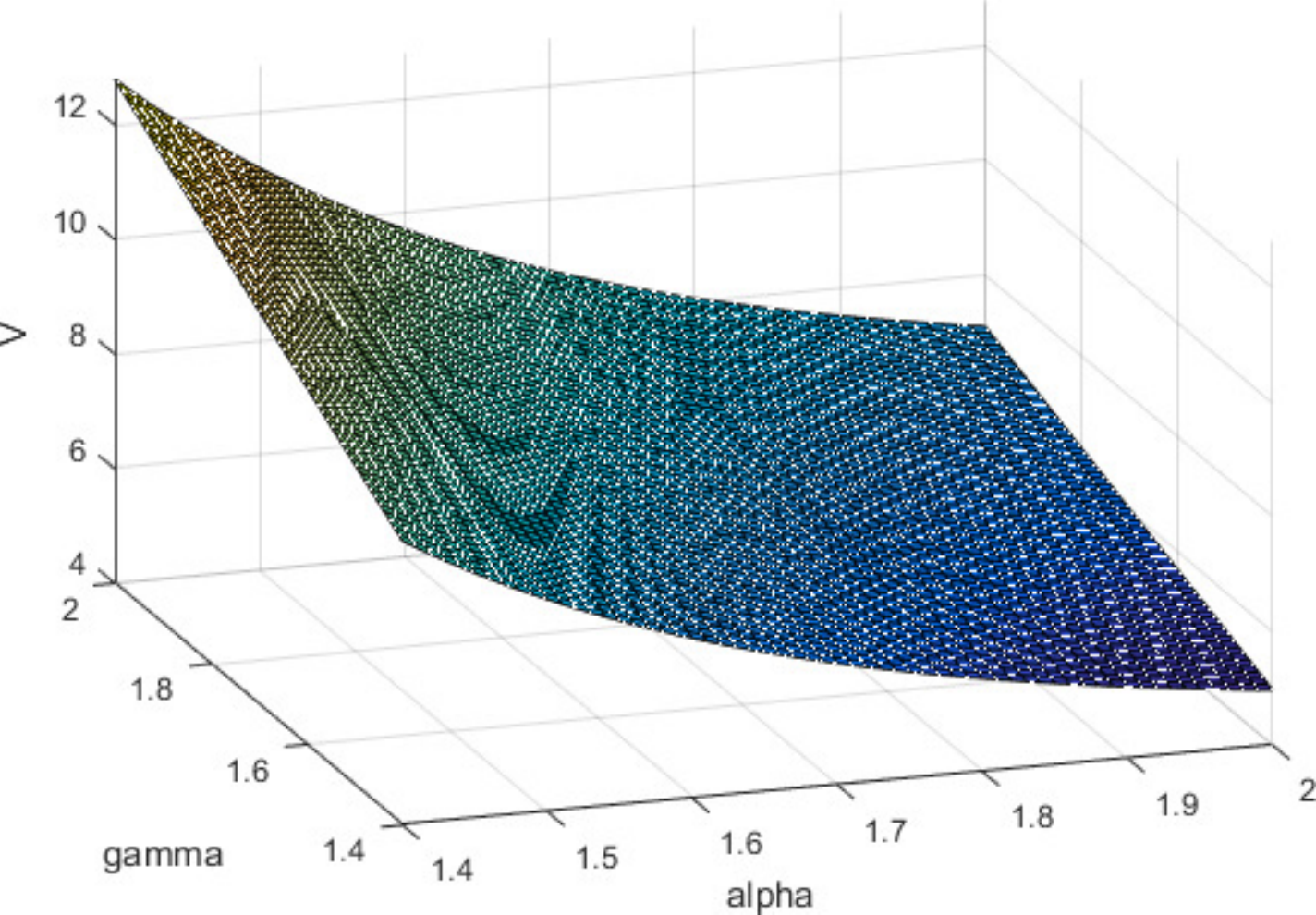}
\caption{\label{v_variation}The 2D normalisation parameter, $V$ varies smoothly over the relevant range of the parameter space defined by $\alpha$ and $\gamma$ and is therefore interpolated as required within the dose calculation engine.}
\end{figure}

To provide radial symmetry of the beam, a 2D normalisation is required when computing the lateral profile in a plane of distance $z$ into a medium. If $S_{z}(x;\alpha,\gamma)$ is the value of the stable distribution that describes the 1D beam profile at a distance $x$ from the central axis, the 2D beam profile is described by:
\begin{equation}
L ~=~ \frac{1}{V}~S_{z}(r;\alpha,\gamma)
\end{equation}
where, $r$ is the distance from the pencil beam central axis and $\alpha$ and $\gamma$ are the parametrization at depth $z$. $V$ is the normalisation required such at the volume under the 2D distribution is unity and is calculated using the shell formula  as:
\begin{equation}
V ~=~ 2 \pi \int_0^\infty x~S(x;\alpha,\gamma)dx
\end{equation}
As there is no analytical representation in real space for stable distributions except when $\alpha$ equals specific values, numerical computation of this integral is required to provide the normalisation. To increase efficiency, the integral $V$ is pre--calculated for each combination of $\alpha$ and $\gamma$ in the discrete beam parametrization and is interpolated as required within the dose calculation engine. Figure \ref{v_variation} shows that $V$ varies smoothly within the relevant range of parameter space.
\par
The parameters required to fully characterise the Knoxville beam at all energies were implemented in \mbox{matRad}: namely, $\alpha$, $\gamma$ and the integrated dose at a distance $z$ along the beam path. This implementation allows complex treatment plans to be generated.

\section{Results}
\subsection{Simulations}
\begin{figure}[h!]
\centering
\subfigure[At Depth 20cm beam: 226.08MeV]{\includegraphics[width=0.9\columnwidth]{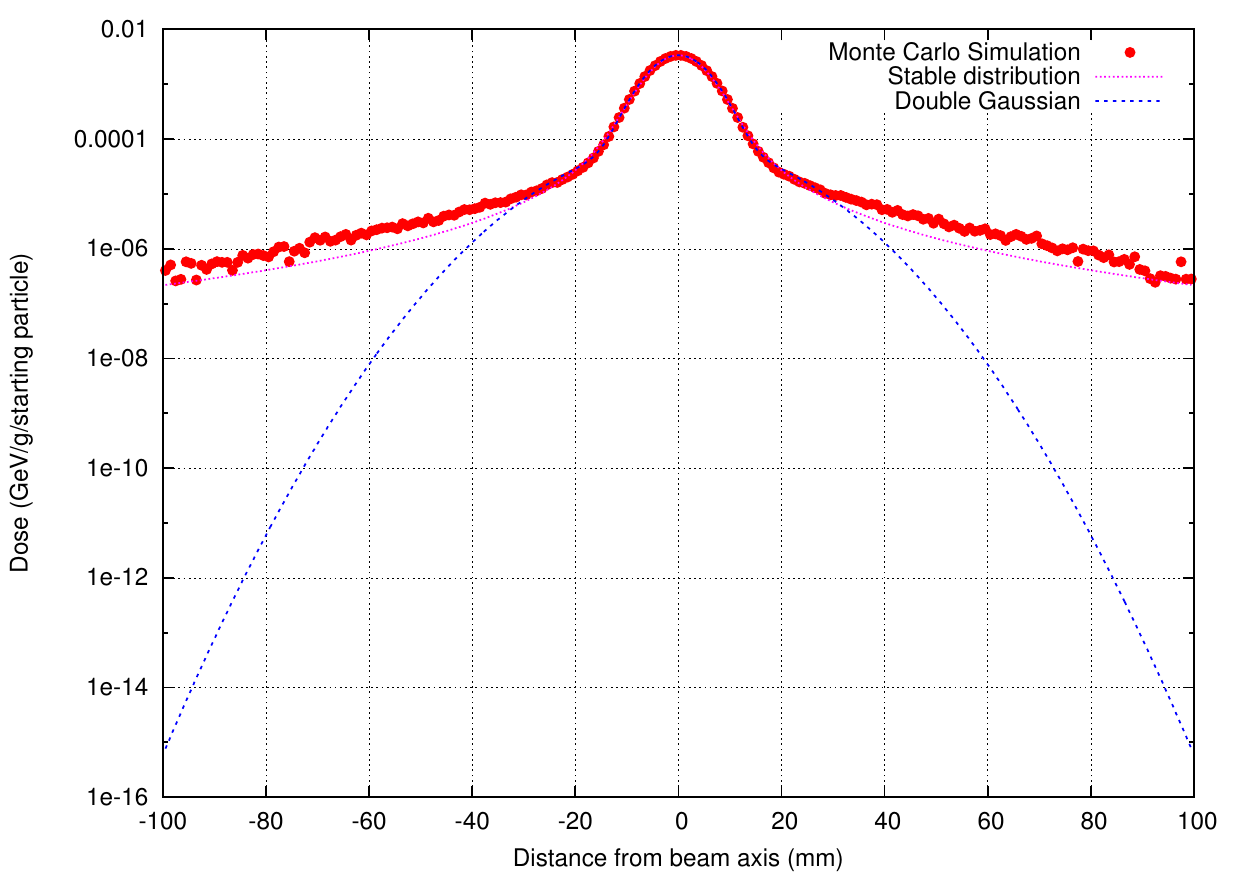}}
\subfigure[Evolution of $\chi^2$--value]{\includegraphics[width=0.9\columnwidth]{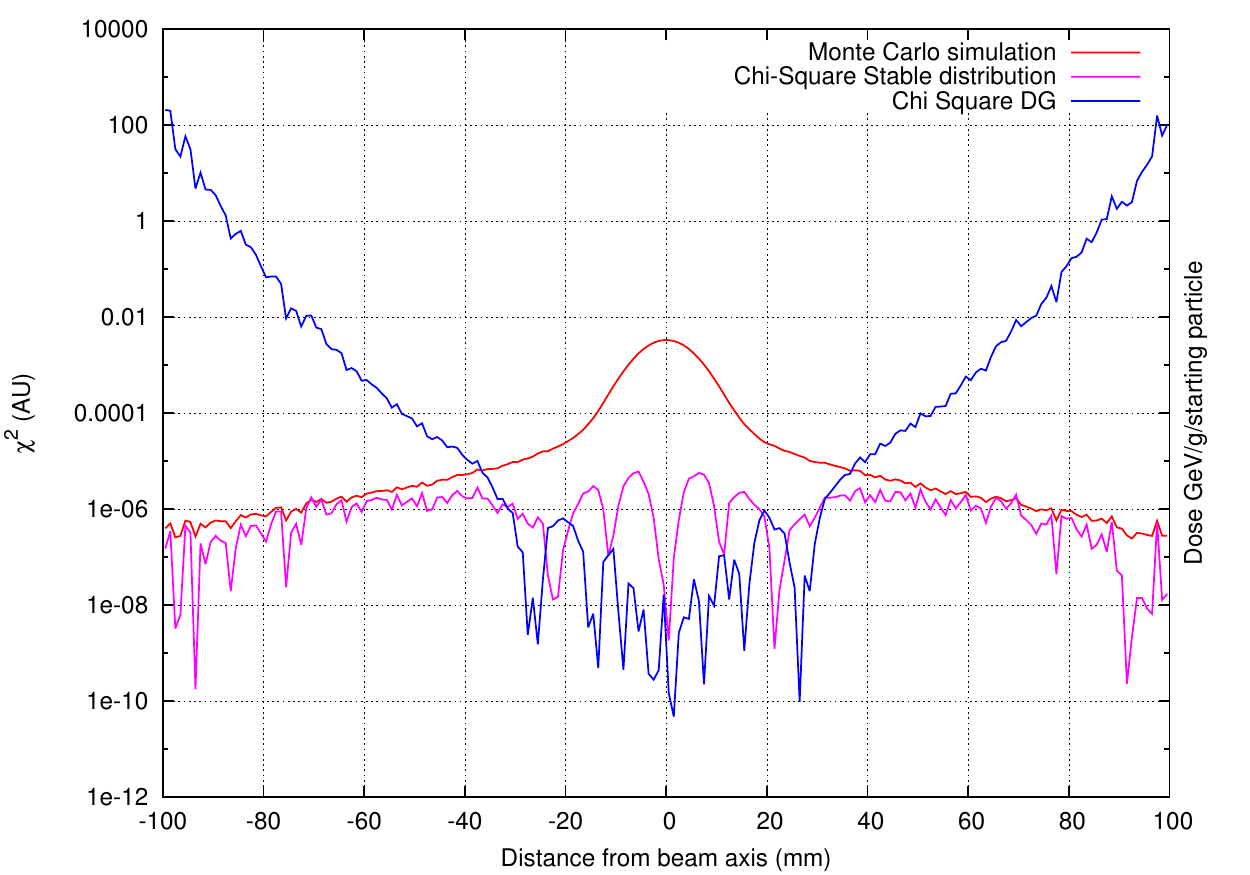}}
\caption{\label{fit1} Quantifying the goodness--of--fit for using a double gaussian and a stable distribution approach.}
\end{figure}
Figure \ref{fit1}a) shows the lateral dose deposition at a depth of 20 cm for a proton pencil beam with nominal energy of 226.08 MeV with two parameterizations: the appropriate stable distribution and a double Gaussian.
From visual inspection of the graph, it is clear that the stable distribution provides a better fit to the beam profile at this energy and depth. The fit of each parameterization is quantified by calculating the $\chi^2$-values, yielding $\chi^2_N = 340.18$ for a single Gaussian distribution (not plotted), $\chi^2_S = 0.00025$ for the stable distribution, and $\chi^2_S = 0.00046$ for a double Gaussian. This corresponds to a probability of 0, 1, and 1, for the lateral dose distributions to be represented by the respective fits.
\par
Figure \ref{fit1}b) plots the $\chi^2_S$--value at each calculated lateral point in the beam profile where a value of 0 is a perfect fit at that point. This graph shows that each of the three distributions provides an adequate representation of the dose close to the centre of the beam. For the Gaussian distribution, the $\chi^2$--values for each point quickly deviates substantially, showing that this distribution does not represent the range behaviour adequately beyond $\sim$10 mm. The double-Gaussian fit provides a good estimation of the beam profile to a distance of $\sim$60 mm, however it becomes clear that there is a systematic underestimation of the dose faraway which increases the $\chi^2$--value. Furthermore, this parameterization requires the most variables to describe the system. The stable distribution provides the best fit of the profile of a proton pencil beam at this energy and depth. 

\par The effectiveness of these systems increases the number of variables needed to describe the system and depends on the region of interest (i.e. the size of the region taken into account to measure the tail contributions).

\subsection{Parameterization}
\begin{figure}[h!]
\centering
\subfigure[$\alpha (z)$]{\includegraphics[width=0.31\textwidth]{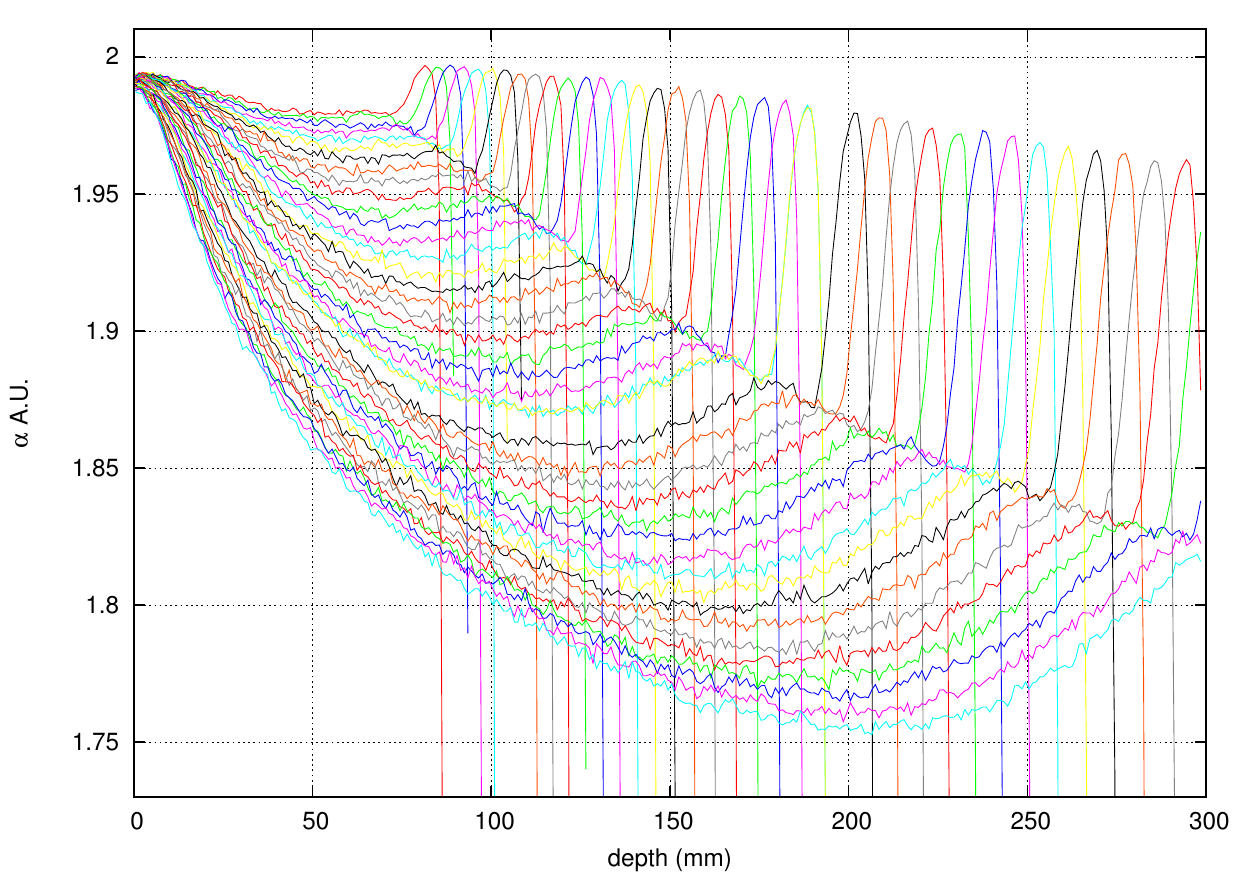}}
\subfigure[$\gamma (z)$]{\includegraphics[width=0.31\textwidth]{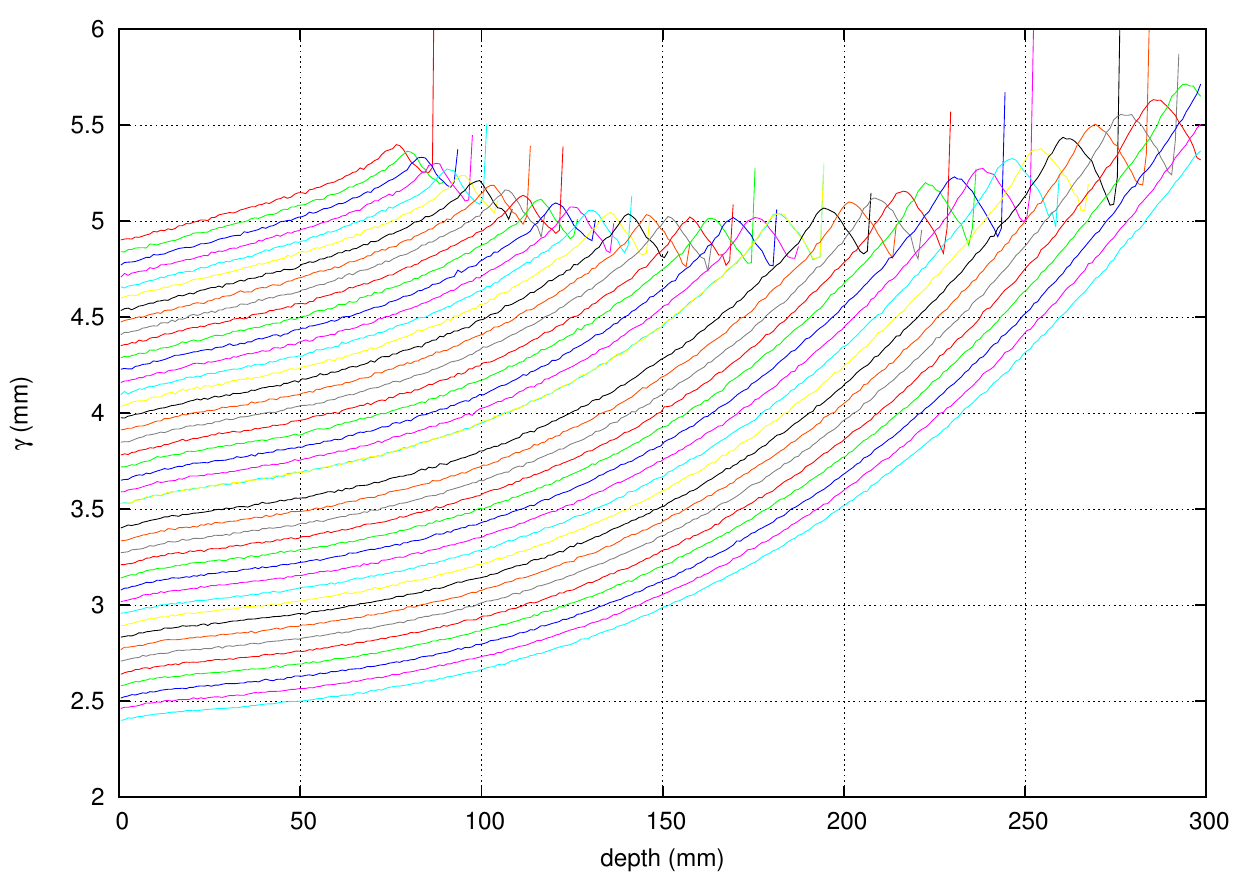}}
\subfigure[$I_D(z)$]{\includegraphics[width=0.31\textwidth]{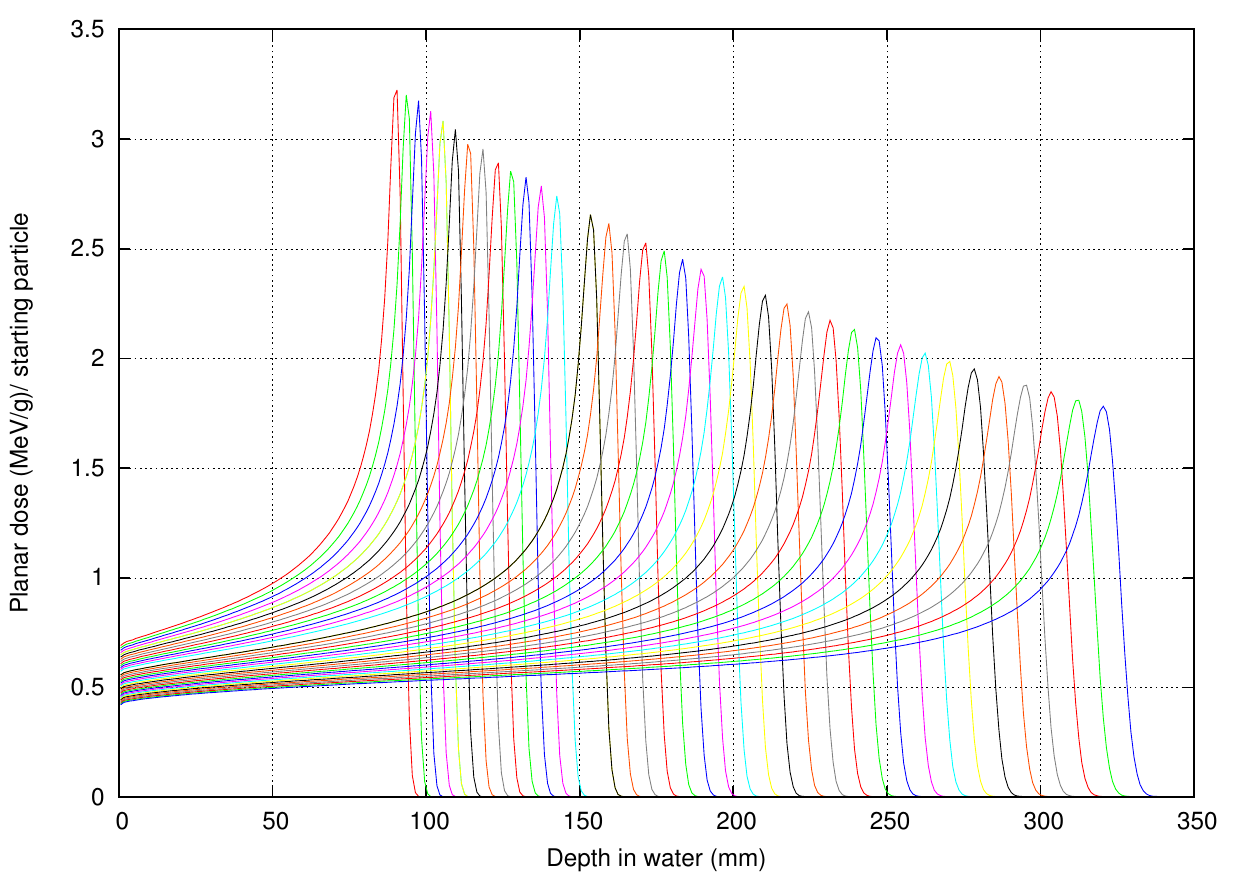}}
\caption{\label{parametrization} From top to bottom: The evolution of $\alpha$, $\gamma$, and integrated dose, note that the graph for each energy has the same general shape}
\end{figure}

The behaviour of a proton pencil beam, as commissioned at the ProVision facility can be parametrised at a given depth and for a specific nominal energy using two parameters from the stable distribution fit: $\alpha$, describing the tail of the distribution and $\gamma$ providing the width. These parameters provide a normalised distribution. A final parameter is the integral dose $I_D$ deposited at that depth (Fig. \ref{parametrization}c) ). The $\alpha$ parameter reflects the increased contribution of interactions with longer range, most likely from scattered protons. The contribution diminishes due to two factors: 1) The decrease of generated secondary protons due to the lower energy of the primary protons, and 2) the decrease in energy of the secondary protons. 

\subsection{Interpolation of data}
\begin{figure}[h!]
\centering
\includegraphics[width=0.45\textwidth]{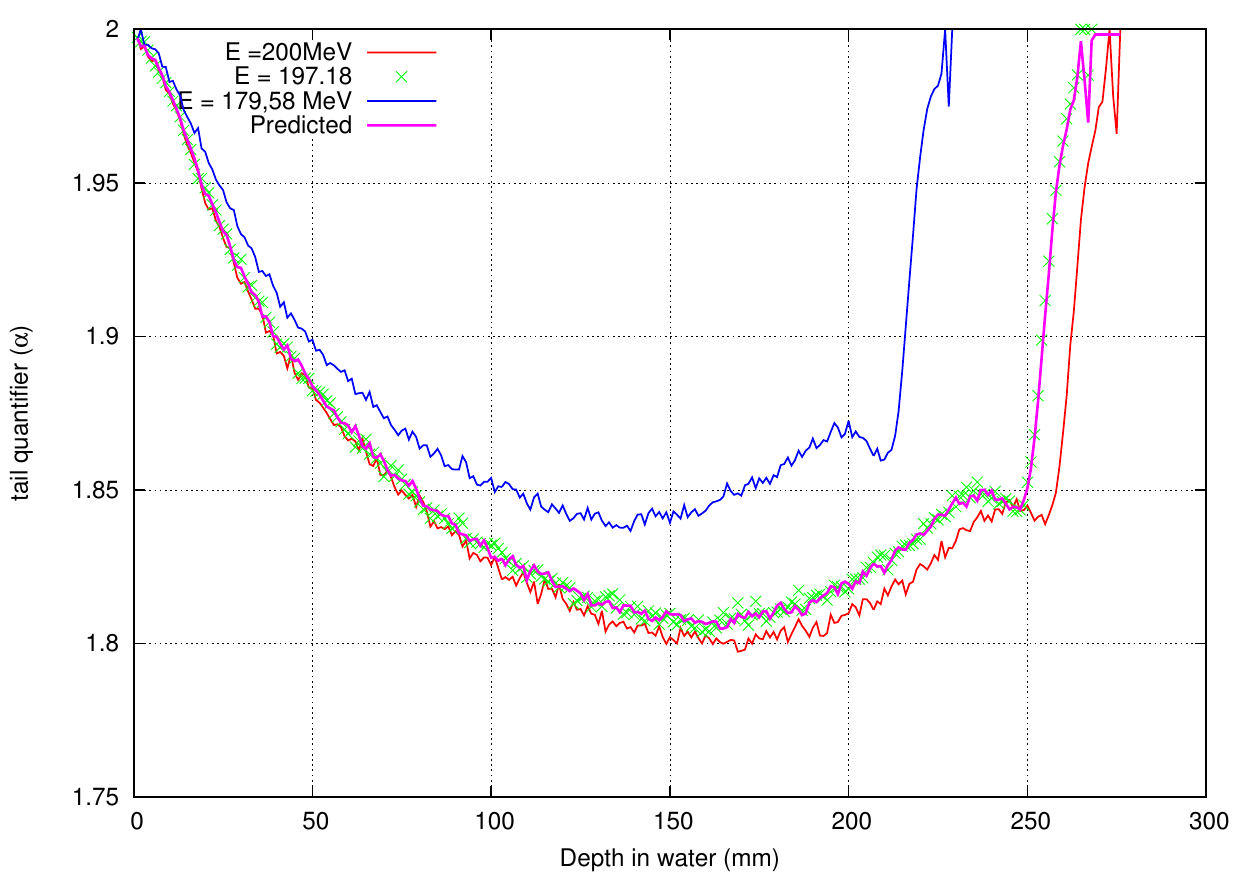}\\
\includegraphics[width=0.45\textwidth]{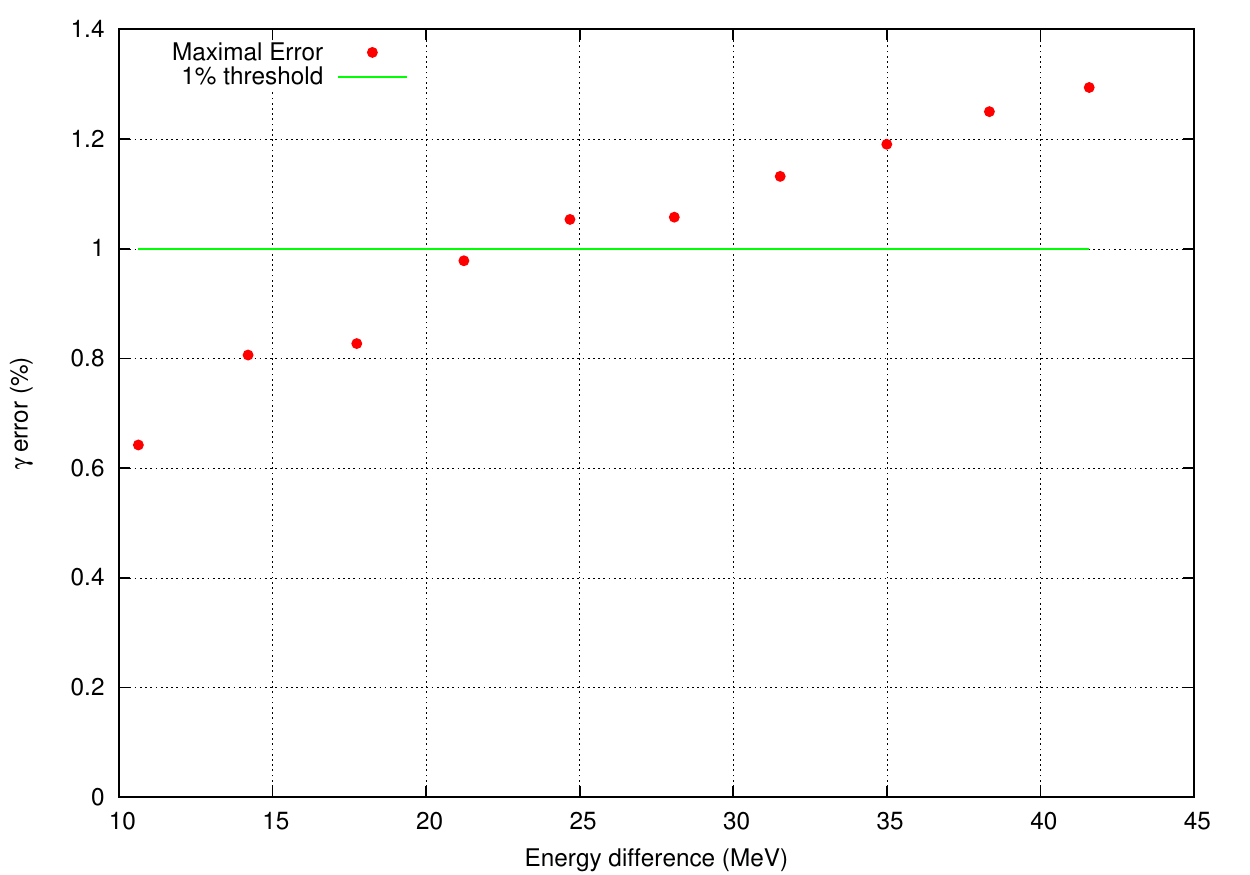}
\includegraphics[width=0.45\textwidth]{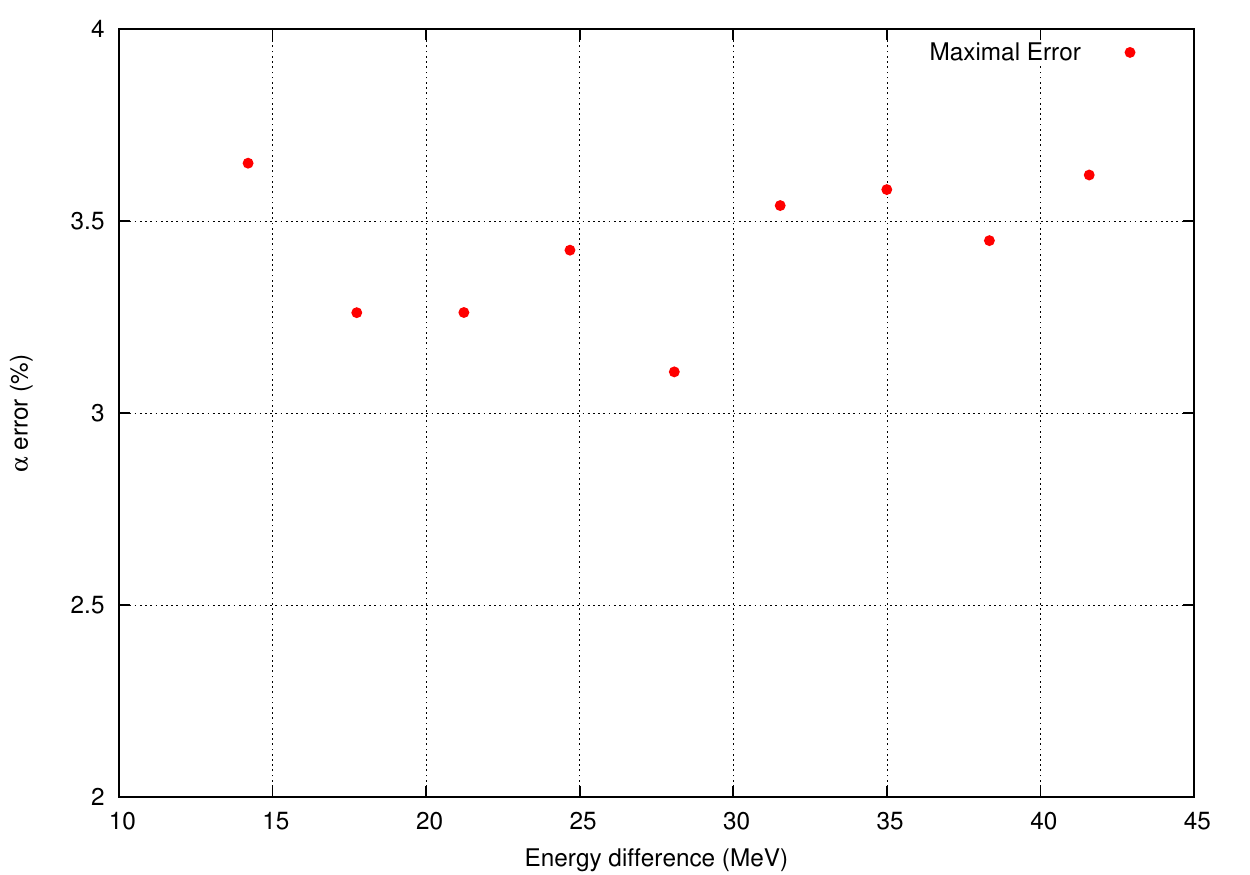}
\caption{\label{interpol} Top figure: Predicting the next $\alpha$--graph using two outer source curves. Left bottom: Variation of $\gamma$ is the limiting parameter, 1\% error equates and error of 0.01 mm in estimation of the width. Right Bottom: The $\alpha$ parameter does not change greatly as it is dominated by the curve noise.}
\end{figure}

Figure \ref{interpol} shows the methodology interpolating the data from two energies to generate data for a third beamlet. In the remaining figures we calculate the maximal error of the parametric representation. Due to the non--linearity of the parameters' behavior as a function of energy we expect that linear interpolation is useful only in a limited energy range. Indeed, Figures \ref{interpol}b) and c), show that the parameter $\gamma$ is most sensitive and increases deviates to more than 1\% if the interpolated energies are more than 20MeV apart in nominal energy. The $\alpha$--parameter is not very sensitive to interpolating distance as the curve is relatively noisy. 

\subsection{matRad Implementation}
The alpha-stable parametrization has been successfully incorporated into the matRad open source treatment planning system. Calculation of alpha-stable distribution is performed using either a fast, parallel C/C++ library \emph{libstable}\cite{Julian-Moreno2016}, if available, or a native MATLAB implementation in other situations\cite{veillette2012stbl}.

\par
Using the C/C++ library, calculation of a complete pencil beam on a 200x200x200 3 mm$^3$ cube takes 20-45 seconds, depending on beam energy, on a Intel Xeon E5-2670 based workstation with ten 2.5 GHz cores.

\par
\begin{figure}[h!]
\centering
\includegraphics[width=1.0\columnwidth]{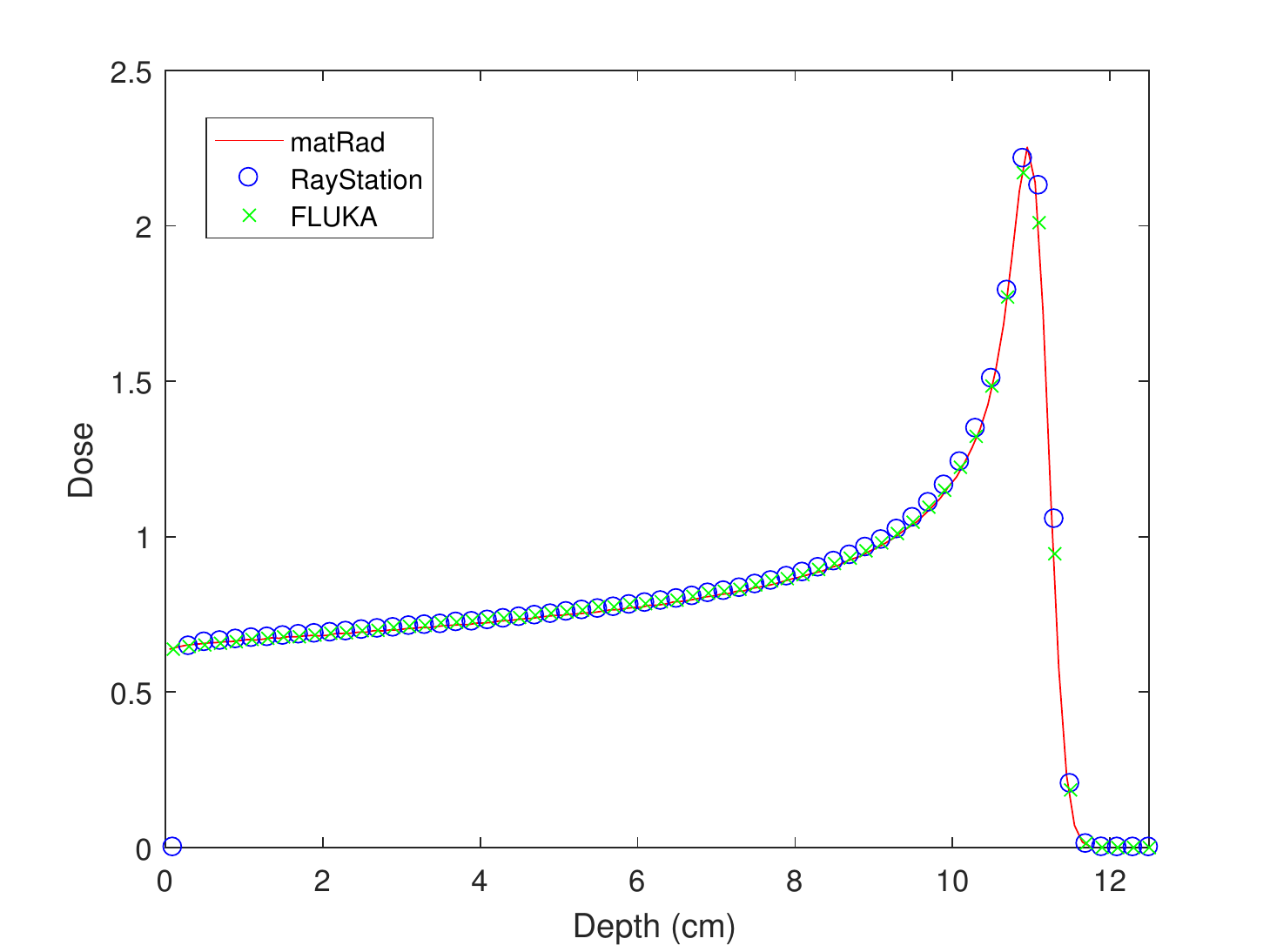}
\includegraphics[width=1.0\columnwidth]{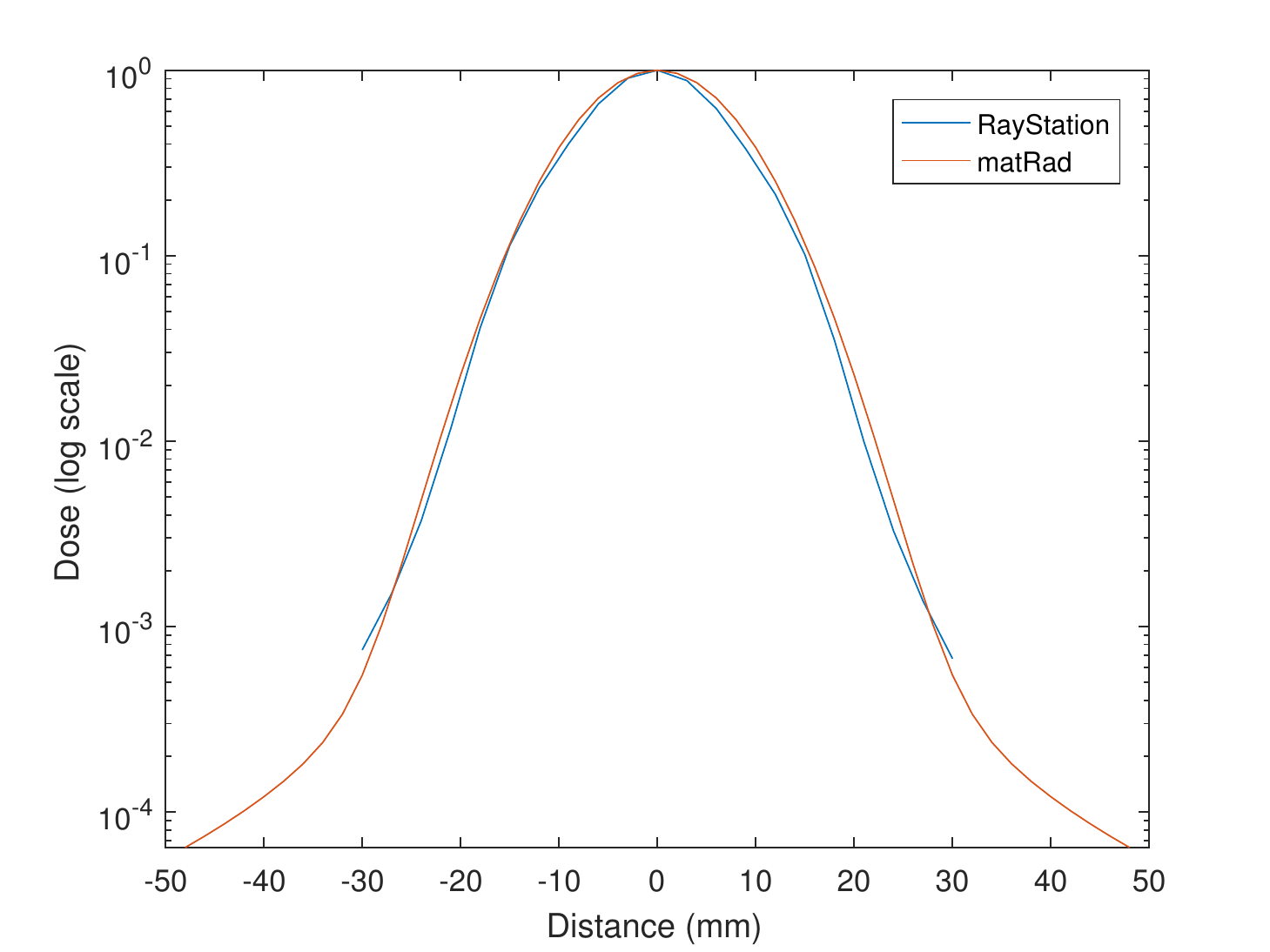}
\caption{\label{raystation_comparison}Comparison of matRad and RayStation calculations of a 120 MeV proton pencil beam dose distribution.}
\end{figure}

The dose distribution from a single proton beam spot of 120 MeV was calculated onto a homogeneous water phantom in matRad, RayStation and FLUKA. The depth dose curve and beam profile across the Bragg Peak are shown in Figure \ref{raystation_comparison}.

\par
\begin{figure}[h!]
\centering
\includegraphics[width=0.8\columnwidth]{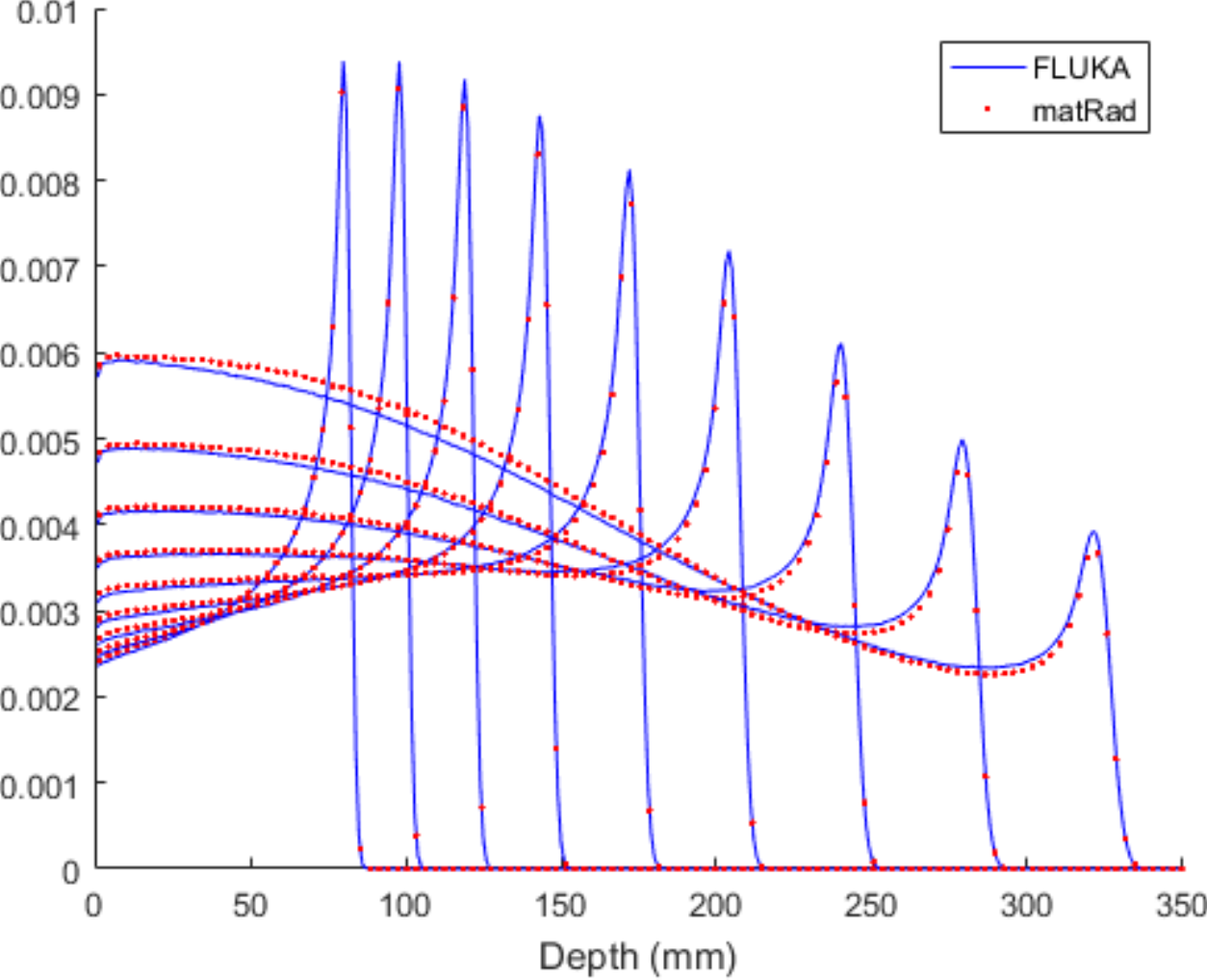}
\caption{\label{matRad_pdd}Original FLUKA simulated central axis beam dose depositions for selected energies (blue) and re-calculated in matRad (red).}
\end{figure}

Figure \ref{matRad_pdd} shows the central axis depth-dose depositions for selected energies from 100.32 to 226.08 MeV form the FLUKA simulations compared to the distributions re-calculated using matRad.
\par
 The matRad calculated dose distributions use a different grid size and spacing to the FLUKA data, specifically 200x200x200 3 mm$^3$ voxels, demonstrating that appropriate lookups as well as interpolation of data are being performed. The small differences seen are due to the fitting of the alpha-stable distributions to FLUKA data.

\section{Discussion}
The use of stable distributions provides a way of calculating the dose in a medium in a scanned pencil beam proton therapy machine that lends itself to implementation on GPU type architectures. The calculation in the Fourier space can be done fast and the libraries are available to do this fast on such processors. Alternatively, it is possible to directly estimate the integral provided by the inverse transform yielding:
\begin{equation}
f(x) ~=~\int_0^\infty \exp(- |\gamma t|^\alpha)\cos(xt)dt
\end{equation}
This can be numerically evaluated using a Gaussian Quadrature method, which is computationally faster than a fast Fourier transform\cite{belov2005computation}. Although Monte Carlo simulation type dose calculators are becoming increasingly available, the use of an analytical alternative is interesting if exhaustive searches in treatment plans are being used. 
\par
Providing a parametrization of this type reduces the number of parameters to a more manageable level, allowing a better insight in the physics of proton therapy planning using scanned pencil beams. It becomes clear, for instance, that the scattering properties of combined scanned beams are different depending on the depth of the treated volume, and therefore different dose characteristics and maybe even biological effects can be expected. This because there might be variations of LET depending on the contributions of the halo at various depths. It also provides a method to describe issues like changes in medium in terms of the used parameters. In a forthcoming study we have already established that not all parameters behave in the same manner as a function of depth combined with changes in material (data not shown).
\par 
In this current study we considered the pencil beam to be isotropic. In practise it is possible that that is not the case depending on the geometric properties of the machine used to generate the pencil beams. 
For instance, many proton therapy facilities will use spatially sequential magnets to bend the beams in the directions perpendicular to the beam axis in two perpendicular directions to each other. This
results in an ellipsoid spot size due to a different virtual source positions. The implication is that we have to find a way to combine different stable distributions. In the case of the normal distribution this is well understood, i.e. combining the variations depending on the mixing angle. Combining generalized stable distributions is less straightforward, but still fairly trivial in the case where the $\alpha$ parameter is constant. Investigating Eq. \ref{cf_short} shows that the combination of two distributions with the same $\alpha$ and scale parameters $\gamma_1$, and $\gamma_2$ yields a new stable distribution with scale parameter $\gamma$:
\begin{equation}
    \gamma = (\gamma_1^\alpha + \gamma_2^\alpha)^{\frac{1}{\alpha}}
\end{equation}
Fortunately, we have seen that the parameter $\alpha$ depends only on the amount of material that has been passed. As a result the value of $\alpha$ is the same in every direction of the plane. Combining stable distributions with different $\alpha$ is not straightforward because as far as we know the resulting distribution is not stable and is still an area of mathematical research.
\par 
We have also limited this study to that of symmetric zero--centered pencil beams. While the zero--centering is easily resolved by a well chosen coordinate transformation, the asymmetry of a pencil beam is not resolvable in an easy way. Indeed, in some cases the treatment beams are not symmetric, specifically if collimation is used and pencil beams near the collimator jaws need to be considered. In that case the parameter $\beta$ is not zero and the full expression as outlined in Eq. \ref{cf_long} needs to be evaluated. This is subject of further research by our group.

\par
In theory, the methodology we have shown here could be extended to other charged particles and photons. This is an area of further research by our group.

\section{Conclusion}
We have demonstrated that alpha--stable distributions are suited to describe charged particle pencil beams in a medium because they provide an accurate and efficient parameterization. We have shown how this parameterization of the pencil beam allows dose distributions from intermediate energies to be interpolated through intermediate morphing. Furthermore, we have implemented the alpha-stable parameterization into a treatment planning system.
\section{Acknowledgements}
Frank Van den Heuvel, Francesca Fiorini, and Ben George greatfully acknowledge core support by Cancer Research UK (CRUK) and the Medical Research Council (MRC). 
\section{Contributions}
Frank Van den Heuvel: Concept, fitting the stable distributions and editing the article, Francesca Fiorini: Monte Carlo simulations and co--writing the article, Niek Schreuder: Measurements to verify pencil beam Monte Carlo calculations and co--writing the article, Ben George: Implementation of the algorithm in matRad and co--writing the article

\bibliographystyle{apsrev4-1.bst}

%

\clearpage
\section*{Appendix A}
In this appendix we outline the notion of stable distributions, provide some definitions and show that the characteristic representation parameterizing the quantities ($\alpha$ and $\gamma$) indeed represents all stable distributions. The text is extensively based on the treatise by Uchaikin and Zolotarev and is provided as a synthesis and guideline rather than an original scientific contribution, the original work is much more extensive and dense \cite{1999_stable}.
\subsection{Defining stable distributions}
We start out by quoting the law of large numbers which states that the difference between the estimated mean of a sample from a random variable tends to the mean of the distribution when enough samples are taken. It is best known in the form as proposed by Bernouilli in the 18th century:
\begin{thm}[law of large numbers --- Jacob Bernoulli]
    Let $X_1$, $X_2$,\ldots $X_n$ be independent, identically distributed random variables with mean $\mu_n = \frac{1}{n}\sum_{i=1}^n X_i$ then:
    \begin{equation}
        P\lbrace |\mu_n - p| \geq \epsilon\rbrace \rightarrow 0 , n \rightarrow \infty
    \end{equation}
    Since
    \begin{equation}
        \epsilon_n = \mu_n - p \overset{P}{\rightarrow}\infty
    \end{equation}
    i.e. it converges to zero in probability as $n \rightarrow \infty$, which provides the reformulation of Bernoulli's law of large numbers:
    \begin{equation}
        \mu_n=p+\epsilon_n, n \geq 1.
    \end{equation}
\end{thm}
A more sophisticated approach considers the random variables as functions on an interval [0,1]  where  $\omega$ is an instantiation of the  experiment yielding 0 or 1. In that case the strong law of large numbers can be replaced by a weaker version:
\begin{equation}
    \mu_n = \sum_{i=1}^n X_i(\omega)  
\end{equation}
which then becomes a degenerate function if infinite samples are taken, but more importantly, before reaching the degenerate condition the sum tends to the normal distribution, which is the classical form of the the central limit theorem. 
    \begin{thm}[central limit theorem --- Moivre--Laplace] Let $X_1$, $X_2$,\ldots be independent, identically distributed random variables with mean $\mu$ and variance $\sigma^2 < \infty$. Then as $n\rightarrow \infty$,
    \begin{equation}
        P\left\lbrace \frac{\sum_{i=1}^{n}X_i - n\mu}{\sigma \sqrt{n}}< x \right\rbrace \Rightarrow \phi(x) = \int_{-\infty}^x p^G(x)dx
    \end{equation}
    Where
    \begin{equation}
        p^G(x) = \frac{1}{\sqrt{2\pi}}e^{-x^2/2}\notag
    \end{equation}
\end{thm}
We also provide the notion of {\em equivalent} distributions $X$ and $Y$:
\begin{equation}\label{equival}
    X\overset{d}{=}Y \iff p_X(x) = p_Y(x)
\end{equation}
As well as the notion of similar distributions which provides the possibility of introducing a linear transformation of the given distribution.
\begin{equation}\label{similar}
    X\overset{s}{=}Y \iff Y \overset{d}{=} a+bX
\end{equation}
Using expression \ref{equival} and \ref{similar} it is clear that for similar distributions $X$ and $Y$ on an infinitesimal interval $dx$: 
\begin{equation}
    p_Y(x)dx = p_{a+bX}(x)dx = p_X\left(\frac{x-a}{b}\right)\frac{dx}{b}
\end{equation}
Therefore, the same applies to the distribution functions (cumulative of the density function):
\begin{equation}
    F_{a+bX}(x) = F_X\left(\frac{x-a}{b}\right)
\end{equation}
For example the normal distribution $p^G(x;0,1)=\frac{1}{\sqrt{2\pi}}\exp(-x^2/2)$ provides the following expression:
\begin{equation}\label{scaling}
    p^G(x;a,\sigma)= \frac{1}{\sigma}p^G(\frac{x-a}{\sigma})
\end{equation}
An interesting property arises when, instead of looking at the variables themselves, we now investigate how sums of these variables (summands) behave. If we have two normal distributions $Y_1$ and $Y_2$ with variances $\sigma_1$ and $\sigma_2$ then it is easy to see that, using expression \ref{scaling}, we get:
\begin{equation}
    \sigma _1 Y^G_1 + \sigma_2 Y^G_2 = \sqrt{\sigma_1^2 + \sigma_2^2}\times Y^G
\end{equation}
By setting $\sigma_1 = \sigma_2 =1$ and with an arbitrary summands, $n$, we obtain a well known result:
\begin{equation}
    \sum_{i=1}^n Y^G_i \overset{d}{=} \sqrt{n}\times Y^G
\end{equation}
Or more interestingly expressed as:
\begin{equation}
    \sum_{i=1}^n Y^G_i \overset{s}{=} Y^G, a =0 , b= \sqrt{n}
\end{equation}
It is the generalization of this property that leads to the notion of stable distributions by allowing arbitrary values of $a$ and $b$.  
\begin{definit}
    A random variable $Y$ is stable if and only if for any arbitrary constants $b'$ and $b''$ there exist constants $a$ and $b$ such that:
    \begin{equation}
        b'Y_1 + b''Y_2 = a + bY
    \end{equation}
\end{definit}
Which then leads to the stable law in the same form as the central limit theorem, as shown by the proof below. 
\begin{thm}[L\'evy]
    Let $X_1$, $X_2$,\ldots be independent, identically distributed random variables, and let there exist constants $b_n > 0$  and $a_n$ such that
    \begin{equation}\label{CLT_stable}
        P\left\lbrace \frac{\sum_{i=1}^{n}X_i - a_n}{b_n}< x \right\rbrace \Rightarrow G(x), n\rightarrow \infty
    \end{equation}
    for some function $G(x)$ which is not degenerate\footnote{By not degenerate we imply that: $G(x)$ is not a step function}. Then $G(x)$ follows the stable law.
\end{thm}
$P$ represents the given probability at the value $x$. While $G(x)$ is the
cumulative probability density function of the distribution.

\subsection{$\alpha,\beta$ representation}
In this section, we show how we move from the expression in \ref{CLT_stable} to the parameterization that we have been using denoting the type of stable distribution based on the single parameter $\alpha$. Before we move on, we narrow the definition of stable distributions to that of {\em strictly} stable distribution by setting $a_n=0$, thus:
\begin{equation}\label{strictly}
    S_n = \sum_{i=1}^n Y_i \overset{d}{=} b_nY.
\end{equation}
Defining $S_n$ the distribution of the sum. By calculating the variance of the distribution \ref{strictly} we obtain:
\begin{equation}
    n\times {\rm var}(Y) = b_n^2\times {\rm var}(Y)
\end{equation}
If ${\rm var}(Y) \neq 0$ and ${\rm var}(Y) < \infty$ then there is only one possibility
\begin{equation}
    b_n \equiv b^G_n = n^{1/2}
\end{equation}
Which reverts to the result obtained for the normal distribution.

\par
With the notion of summands, we can now use them to further extend the properties to the general case of summing strictly stable random variables. 
\begin{align}
    Y_1 + Y_2 & \overset{d}{=} b_2 X\\\notag
    Y_1 + Y_2 + Y_3 &\overset{d}{=} b_3 X\\\notag
    Y_1 + Y_2 +Y_3 + Y_4 &\overset{d}{=} b_4 X\notag
\end{align}
We now choose to limit ourselves to summands with $2^k$ terms
\begin{align}\label{2ksum}
    Y_1 + Y_2 & \overset{d}{=} b_2 X\\\notag
    Y_1 + Y_2 +Y_3 + Y_4 &\overset{d}{=} b_4 X\\\notag
    Y_1 + Y_2 +Y_3 + Y_4 + Y_5 + Y_6 +Y_7 + Y_8&\overset{d}{=} b_8 X\\\notag
    \ldots&\\\notag
    Y_1 + Y_2 + \ldots + Y_{2^{k-1}} + Y_{2^k} &\overset{d}{=} b_{2^k} X\\ \notag
    \ldots&\\\notag
\end{align}
Keeping in mind that $X_1 +X_2 \overset{d}{=} X_3 + X_4$, which we can generalize to any pair, we can rewrite the expression in (\ref{2ksum})
to read:
\begin{align}\label{2ksum2}
    Y_1 + Y_2 & \overset{d}{=} b_2 X\\\notag
    Y_1 + Y_2 +Y_3 + Y_4 &\overset{d}{=} {b_2}^2 X\\\notag
    Y_1 + Y_2 +Y_3 + Y_4 + Y_5 + Y_6 +Y_7 + Y_8&\overset{d}{=} {b_2}^3 X\\\notag
    \ldots&\\\notag
    Y_1 + Y_2 + \ldots + Y_{2^{k-1}} + Y_{2^k} &\overset{d}{=} {b_2}^k X\\ \notag
    \ldots&\\\notag
\end{align}
Or in much shorter notation
\begin{equation}
    S_{2^k} = b_{2^k}Y =  b_2^kY
\end{equation}
and using the expression in defining strictly stable random variables\ref{strictly}: $S_n = b_n Y$, we obtain:
\begin{equation}
    b_n = b_2^k = b_2^{(\ln n)/\ln 2}     
\end{equation}
transforming into
\begin{equation}
    \ln b_n = [(\ln n) / \ln 2] \ln b_2 = \ln n^{(\ln b_2)/\ln 2}
\end{equation}
or 
\begin{equation}\label{2k}
    b_n =  n^{(\ln b_2)/\ln 2} = n^{1/\alpha_2}
\end{equation}
We can now repeat this process with $3^k$, $4^k$ and using induction $m^k$, yielding $\alpha_3$, $\alpha_4$, and $\alpha_m$, bunching respectively 3, 4 and $m$ summands. In general the following expression is valid:
\begin{equation}
    b_n=n^{1/\alpha_m},~\alpha_m= (\ln m)/\ln b_m
\end{equation}
For arbitrary values of $m$.

\par
If we now set $m=4$ we get 
\begin{equation*}
    \alpha_4 = (\ln 4)/\ln b_4
\end{equation*}
On the other hand, selecting $k = 2$ in expression (\ref{2k}), yields
\begin{equation*}
    \ln b_4 = {1/\alpha_2}\ln 4
\end{equation*}

\par
From these last two formula we conclude that $\alpha_2 = \alpha_4$. By induction we see that there is a single $\alpha$ for all these stable distributions, and that the scaling factors $b_n$ follow the following law:
\begin{equation}
    b_n = n^{1/\alpha}
\end{equation}

\subsection{Characteristic function of symmetrical zero--centered stable distribution}
The characteristic function, c.f., of a distribution can be defined as the Fourier transform of the probability density function, p.d.f., of that distribution. Let $p_X(x)$ be a p.d.f. of a set of random variables $X$, then the c.f., $f_p(k)$, is defined as it's expectation value, $e^{\iu kX}$:
\begin{equation}
    f_X(k) = \int_{-\infty}^{\infty} p_X(x) e^{\iu k} dx
\end{equation}
From this definition some properties follow immediately:
\begin{enumerate}
    \item $f_X(0) =1$
    \item $f_{a+bX} = e^{\iu ka}f_X(bk)$
    \item $f^*_X(k) = f_X(-k) = f_{-X}(k)$, where $*$ denotes the complex conjugate.
    \item If $X$ is symmetric about zero, $$X \overset{d}{=} -X $$ 
    \item If $\mathbf{E}|X|^n,~n\geq1$ then the continuous $n$th derivative of the c.f. exists and: $$f^{(n)}(0) = \iu^n\mathbf{E}X^n$$
    \item if $S_n$ is the sum of independent random variables $X_1$, $X_1$, \ldots , $X_n$, then: $$f_{S_n}(k) = f_{X_1}(k)\ldots f_{X_n}(k)$$
    \item any $f_X(k)$ is a uniformly continuous function. 
\end{enumerate}
In order to progress further, we invoke the inversion theorem to be able to find the distribution function given the c.f.
\begin{thm}{(Inversion theorem)}
    Any distribution function $F(x)$ is uniquely defined by its c.f. $f(k)$. If $a$ and $b$ are some continuity points if $F(x)$, then the inversion formula states that
    \begin{equation}
        F(b) -F(a) = \underset{c\rightarrow\infty}{\lim}\frac{1}{2\pi}\int_{-c}^{c}\frac{e^{-\iu kb} - e^{\iu ka}}{ik}f(k)dk
    \end{equation}
\end{thm}
The principle advantage of using the c.f. is that the c.f. of a sum of independent random variables is equal to the product of the c.f.s of the summands:
\begin{equation}\label{conv_prop}
    f_{X_1+X_2}(k) = f_{X_1}(k)f_{X_2}(k),
\end{equation}
If we take the logarithm (obtaining the second characteristic: $\psi_X(k) = \ln f_X(k)$), we find that:
\begin{equation}
    \psi_{X_1+X_2}(k) = \psi_{X_1}(k) + \psi_{X_2}(k)
\end{equation}
This is important as it allows us to assess the summation of a large number of independent random variables without evaluating multiple integrals. For this reason we cite the continuity theorem:
\begin{thm}{(continuity theorem )}
    Let $f_n(k), n= 1,2,\ldots$ be a sequence of c.f.s and let $F_n(x)$ be a sequence of corresponding distribution functions. If $f_n(k)\rightarrow f(k)$ as $n\rightarrow \infty$, for all $k$ and $f(k)$ is continuous at $k=0$, then $f(k)$ is the c.f. of a cumulative distribution function $F(x)$, and the sequence $F_n(x)$ weakly converges to $F(x)$, $F_n \Rightarrow F$. The inverse is also true: if $F_n \Rightarrow F$ and $F$ is a distribution function, then $f_n(k)\rightarrow f(k)$, where $f(k)$ is the c.f. of the distribution function $F$.
\end{thm}
To gain some insight in how to perform this, we can look at two well known stable distributions to find a way forward. The distributions under consideration are the normal distribution and the Cauchy distribution. The calculation of the characteristic function for these distributions is well known and they also form part of the stable distribution, in the form $q(x;\alpha,\beta)$ representing the stable distribution density:
\begin{description}
    \item[Normal distribution] $q(x;2,0) = \frac{1}{2\sqrt{\pi}}e^{-x^2/4}$
    \item[Cauchy distribution] $q(x;1,0) = \frac{1}{\pi}\frac{1}{1+x^2}$
\end{description}
The characteristic function, $g(k;\alpha,\beta)$ in this notation, can then be calculated in a straightforward manner and can be found in many textbooks:
\begin{description}
    \item[Normal distribution] $g(k;2,0) = e^{-k^2}$
    \item[Cauchy distribution] $g(k;1,0) = e^{-|k|}$
\end{description}
Note that the traditional form of the density function for the normal distribution is slightly different:
\begin{equation}
    p^G(x) =\frac{1}{\sqrt{2\pi}} e^{-x^2/2}
\end{equation}
This observation allows us to generalize for any symmetric distributions. Let $Y_i$ be such a distribution with an arbitrary parameter $\alpha$. We remind that 
\begin{equation}
    \sum_{i=1}^nY_i \overset{d}{=} b_n Y,\quad b_n=n^{1/\alpha}
\end{equation}
Keeping in mind the property as elucidated in equation \ref{conv_prop} 
\begin{equation}
    f_Y^n(k) = f_Y(n^{(1/\alpha)}k)
\end{equation}
using the expression for the second characteristic
\begin{equation}
    n\psi_Y(k) = \psi_Y(n^{(1/\alpha)}k)
\end{equation}
The results obtained earlier when investigating the normal- and the Cauchy distribution lead us to propose a solution of the form
\begin{equation}
    \psi_Y(k) = -c k^\mu,\quad k>0
\end{equation}
This satisfies the previous expression with $\mu = \alpha$ and an arbitrary complex--valued $c$, which we choose to be
\begin{equation}
    c= \lambda[1-\iu c_1],\quad \lambda,c_1, \text{ real numbers}
\end{equation}
To find $\psi_Y(k)$ we use property (3) of the chacaracteristic function, the link between conjugate characteristic function and negative estimates:
\begin{equation}
    \psi_Y(k) = -\lambda[|k|^\alpha-\iu k \omega(k)], \quad -\infty<k<\infty, 0<\alpha\leq 2
\end{equation}
with
\begin{equation}
    \omega(k) = c_1|k|^{\alpha-1}
\end{equation}
Which is a trick to rewrite the equation as $k\times k^{\alpha-1} = k^\alpha$, in such a way that we are explicitly splitting the expression in a real and an imaginary part by a good choice of the constant $c$. We also have not specified what form the function $\omega(k)$ takes, we do know that it will depend on the parameter $\alpha$ as well as provide a measure for the asymmetry of the distribution, should that be present. Later we will attribute that to a parameter $\beta$. The constant $\lambda$ is an arbitrary real number and can serve as a scaling factor, which can be renormalised to $1$, without loss of generality. This implies that the full expression of a stable distributions characteristic function is of the form:
\begin{equation}
g(k;\alpha,\beta) = \exp(-|k|^\alpha + \iu k \omega(k;\alpha,\beta))
\end{equation}
Taking into account that the characteristic function of a symmetric stable function is real--valued due to property (3) of the characteristic functions:
\begin{equation}
    \omega(k;\alpha,\beta) = 0
\end{equation}
and the characteristic function for any stable function becomes:
\begin{equation}
    g(k;\alpha,0,\gamma) = \exp(-|\gamma k|^\alpha)
\end{equation}
With $\gamma$ as a scaling factor.

\end{document}